\documentclass[12pt]{iopart}
\usepackage{graphicx}
\usepackage{iopams}

\newcommand{\ket}[1]{| #1 \rangle}
\newcommand{\bra}[1]{\langle #1 |}
\newcommand{\overleftrightarrow}[1]{\vec{\vec{ #1}} }
\begin{document}
\title{Off-resonant transitions in the collective dynamics of multilevel atomic ensembles}
\author{Yevhen~Miroshnychenko, Klaus~M{\o}lmer}
\address{  Department of Physics and Astronomy\\
  University of
  Aarhus\\
  DK 8000 Aarhus C, Denmark} \date{\today}
\begin{abstract}
We study the contributions of off-resonant transitions to the dynamics of a system of $N$ multilevel atoms sharing one excitation and interacting with the quantized vector electromagnetic field. The Rotating Wave Approximation  significantly simplifies the derivation of the equations of motion describing the collective atomic dynamics, but
it leads to an incorrect expression for the dispersive part of the atom-atom interaction terms. For the case of two-level atoms and a scalar electromagnetic field, it turns out that the atom-atom interaction can be recovered  correctly if integrals over the photon mode frequencies are extended to incorporate negative values. We explicitly derive the atom-atom interaction for multi-level atoms, coupled to the full vector electromagnetic field, and we recover also in this general case the validity of the results obtained by the extension to negative frequencies of the formulas derived with the Rotating Wave Approximation.
\end{abstract}
\maketitle
\section{Introduction}
\label{sec:intro}
The Jaynes-Cummings model of a single two-state atom interacting with a single quantized field mode \cite{JC} constitutes at the same time a corner stone for multiple studies in modern quantum optics and a toy model for fundamental phenomena such as spontaneous and simulated emission, Lamb shifts, dipole- and rotating wave approximations. A multitude of new physical systems allow implementation of the Hamiltonian, sometimes with the roles of fields and atoms replaced by other oscillator and two state systems, while adaption of the theory is needed to deal with novel applications of the field-atom interaction, e.g., for quantum information purposes. In this work we present an analysis of the coherent evolution of a collection of two state atoms and many modes of the quantized radiation field. We consider the case with one single excitation shared by the physical system, and in the spirit of the analytical solution of the unitary Jaynes-Cummings dynamics, we solve for the combined state vector of the atoms and the quantized field.

Since the early work of Dicke \cite{Dicke54}, the problem of collective emission from a sample of identical two-level atoms has received much attention, see e.g. \cite{Friedberg11,Svidzhinsky12} and references therein. The problem is typically treated using the Rotating Wave Approximation (RWA) to mathematically simplify the derivation. A recent series of papers \cite{Friedberg08, Svidzinsky08, Friedberg08a, Scully09b, Svidzinsky10} has discussed the validity of the RWA and, particularly, the importance of effects due to virtual photons in a system of two-level atoms and a scalar light field. The equations of motion without RWA for a vector field interacting with a system of two-level atoms were derived by Milonni \textit{et al.} \cite{Milonni74}, who identified differences between the RWA and the full equations of motion. They also found that the atom-atom interaction terms which were not obtained correctly by the RWA derivation, can be obtained if integrals over photon mode frequencies in the RWA expression are extended to negative values. This artificial extension of the integration range leads to the inclusion of rapidly oscillating terms which, indeed, contrast the argument for applying the RWA in the first place. The artificial, negative energy photon states, however, formally violate energy conservation by the same amount as the far-off resonant, virtual states, which may really influence the atomic evolution. This may explain why they lead to the same dispersive coupling among the atoms.

The equivalence of the two calculations has been demonstrated by going through both calculations and comparing the final results for the two-level atoms with scalar fields. It has not however been proven by a physical argument. Hence it is not clear if a similar simplification holds in the general case of multi-level atoms and vector electromagnetic fields. Although the full equations of motion without the RWA in this general case have been derived by Friedberg \textit{et al.} \cite{Friedberg08b}, see as well \cite{Kiffner10,Wang10}, the question about the \textit{equivalence} of this result to the extended RWA derivation remains open. In this work we explicitly derive the equations following from the usual RWA and from the RWA with the extension to negative frequencies and present the parallel to the full equations. Our results confirm, also for this general case, the validity of the much simpler extended RWA ansatz.

The paper is organized as follows. In Sec. II we derive the general multi-atom collective dynamics within the RWA and within the RWA extended to negative frequencies. In Sec. III we derive the general multi-atom dynamics, including virtual excitation of non-RWA dipole coupled states with one photon and two excited atoms. Sec. IV summarizes and concludes the paper. Evaluation of a few matrix elements and integrals are provided in two appendices.

\section{Rotating Wave Approximation}
\label{sec:RWA}
We consider a system of $N$ atoms located at rest at positions $\vec{r}_j$. Each atom has three degenerate excited states $\ket{e^{-1}}$, $\ket{e^{0}}$ and $\ket{e^{+1}}$ with excitation energy $\hbar \omega_0$ above a single ground state $\ket{g}$ and a dipole moment $d_{eg}$. The three excited states decay with the same rate $\Gamma=\frac{4d_{eg}^2\omega_0^3}{3\hbar c^3}$ and emit photons with the polarization dictated by dipole selection rules.

The Hamiltonian describing the system in the Schr\"odinger picture is \cite{Milonni74,Knight80,Miroshnychenko12}
 \begin{equation}
  \label{eq:H}
  H = H_0+H_{V},
\end{equation}
where the free and interaction Hamiltonians are
 \begin{equation}
  \label{eq:H0}
  H_0 = \sum_{\vec{k}, \lambda}\hbar \omega_k a_{\vec{k} \lambda}^+ a_{\vec{k} \lambda}+\sum_{j=1}^N \sum_{\nu=-1}^1 \hbar \omega_0 \ket{e_j^{\nu}} \bra{e_j^{\nu}}
\end{equation}
and
 \begin{equation}
  \label{eq:HV_general}
  H_V =- \sum_{j=1}^N \vec{D}^j \cdot \vec{E}\left(\vec{r}_j \right).
\end{equation}
Taking the transition dipole moment operator for atom~$j$
 \begin{equation}
  \label{eq:D}
\vec{D}^j=d_{eg}(\vec{\sigma}^j_{eg}+\vec{\sigma}^j_{ge})
\end{equation}
and the electric field operator evaluated at the position of this atom
 \begin{equation}
  \label{eq:E}
\vec{E}\left(\vec{r}_j \right)=i \sum_{\vec{k},\lambda}  \left( \frac{2 \pi \omega_k \hbar}{ V}\right)^{\frac{1}{2}} \vec{\epsilon}_{\vec{k}, \lambda} \left(a_{\vec{k}, \lambda} e^{i \vec{k} \cdot \vec{r}_j}-  a^+_{\vec{k}, \lambda} e^{-i \vec{k} \cdot \vec{r}_j}\right),
\end{equation}
we can write the interaction Hamiltonian in the form
\begin{equation}
\label{eq:HV}
  H_V = -i \sum_{j=1}^N \sum_{\vec{k},\lambda}\hbar g_k \left(\vec{\sigma}_{eg}^j+ \vec{\sigma}_{ge}^j\right) \cdot \left( \vec{\epsilon}_{\vec{k} \lambda} a_{\vec{k} \lambda} e^{i \vec{k}\cdot \vec{r}_j}- \vec{\epsilon}_{\vec{k} \lambda} a_{\vec{k} \lambda}^+ e^{-i \vec{k}\cdot \vec{r}_j} \right).
\end{equation}
Here $a_{\vec{k} \lambda}^+$ denotes the photon creation operator in a mode with wave vector $\vec{k}$, energy $\omega_k=c k$ and polarization along direction $\vec{\epsilon}_{\vec{k} \lambda}$, $\lambda =1,2$. The vacuum coupling constant is $g_k=d_{eg} \left( \frac{2 \pi \omega_k}{\hbar V}\right)^{\frac{1}{2}}$. The dipole operators for atom $j$ are $\vec{\sigma}_{eg}^j=\sum_{\nu=-1}^1 \hat{d}_{\nu g} \ket{e_j^{\nu}}\bra{g_j}$ and $\vec{\sigma}_{ge}^j=\sum_{\nu=-1}^1 \hat{d}_{g \nu} \ket{g_j}\bra{e_j^{\nu}}$
with the complex unit vectors $\hat{d}_{0 g}=(0,0,1)$, $\hat{d}_{-1 g}=\frac{1}{\sqrt{2}}(1,-i,0)$ and $\hat{d}_{+1 g}=\frac{1}{\sqrt{2}}(1,i,0)$. We assume a real polarization basis, $\vec{\epsilon}_{\vec{k} \lambda}$ \cite{Milonni74}.
\begin{figure}[tbp]
  \centering
{\includegraphics{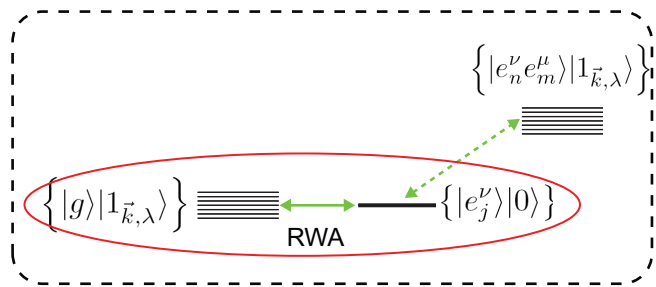}}
  \caption{(Color online) Couplings due to the interaction Hamiltonian $H_V$. The states are arranged to reflect the number of atomic excitations and the relative energy. The system with initially only one atomic excitation is directly coupled to the atomic ground state with one photon emitted and off-resonantly coupled to a state with two atomic excitations and one photon. In the RWA this off-resonant coupling is explicitly neglected.
\label{fig:RWA}}
\end{figure}

In the Rotating Wave Approximation only the terms that conserve the total number of excitations in the atomic ensemble and the quantized field are retained in the interaction Hamiltonian Eq.~(\ref{eq:HV})
\begin{equation}
\label{eq:HVRWA}
  H^{RWA}_V = -i \sum_{j=1}^N \sum_{\vec{k},\lambda}\hbar g_k \left[ \left(\vec{\sigma}_{eg}^j \cdot \vec{\epsilon}_{\vec{k} \lambda}\right) a_{\vec{k} \lambda} e^{i \vec{k}\cdot \vec{r}_j} -  \left(\vec{\sigma}_{ge}^j \cdot\vec{\epsilon}_{\vec{k} \lambda}\right) a_{\vec{k} \lambda}^+ e^{-i \vec{k}\cdot \vec{r}_j} \right].
\end{equation}

The RWA Hamiltonian Eq.~(\ref{eq:HVRWA}) couples only the states connected by the horizontal arrow in Fig.~\ref{fig:RWA}. Therefore, if there is initially only one excitation in the system, the wave function at any later time has the form
\begin{equation}
\label{eq:psiRWA}
 \ket{\psi(t)}=\sum_{j=1}^N \sum_{\nu=-1}^1 \beta_j^{\nu}(t)e^{-i\omega_0 t}\ket{e_j^{\nu}}\ket{0}+\sum_{\vec{k},\lambda} e_{\vec{k} \lambda}(t)e^{-i \omega_k t} \ket{g}\ket{1_{\vec{k} \lambda}}.
\end{equation}
Here we have introduced the following notations: $\ket{g}=\ket{g_1,g_2,...,g_N}$ with no atomic excitations, and $\ket{e_j^{\nu}}=\ket{g_1,...,e_j^{\nu},...,g_N}$ with one excited atom, respectively. $\ket{0}$ and $\ket{1_{\vec{k} \lambda}}$ denote photon states with no photon and with one photon in the $\vec{k}$ mode with the polarization $\lambda$, respectively.

Substitution of Eq.~(\ref{eq:psiRWA}) into the time dependent Schr\"odinger equation with the Hamiltonian from Eqs.~(\ref{eq:H}), (\ref{eq:H0}) and (\ref{eq:HVRWA}) yields

\begin{equation}
  \label{eq:eRWA}
 \dot{e}_{\vec{q} \sigma}=\sum_{j=1}^N \sum_{\nu=-1}^1 \beta_j^{\nu}g_q e^{i((\omega_q-\omega_0)t- \vec{q} \cdot \vec{r}_j)}\left( \hat{d}_{g \nu}\cdot \vec{\epsilon}_{\vec{q} \sigma}\right)
\end{equation}
and
\begin{equation}
  \label{eq:betaRWA}
 \dot{\beta}_{l}^{\eta}=-\sum_{\vec{k}, \lambda} e_{\vec{k} \lambda}g_k e^{-i((\omega_k-\omega_0)t- \vec{k} \cdot \vec{r}_l)}\left( \hat{d}_{\eta g}\cdot \vec{\epsilon}_{\vec{k} \lambda}\right).
\end{equation}
In the Appendix~\ref{sec:AppendixA} we derive explicitly all the matrix elements needed for the analysis.\\

Starting with an atomic excitation, and thus the initial condition $e_{\vec{q} \sigma}(0)=0$, the Equations~(\ref{eq:eRWA})-(\ref{eq:betaRWA}) can be formally integrated. Here we additionally assume the Markovian approximation, i.e., there is no atomic population dynamics faster than the time scale of the atomic decay: $\beta_j^{\nu}(\tau)\approx \beta_j^{\nu}(t)$ for $|t-\tau| \ll \Gamma^{-1}$ \cite{Scully97}. This yields
\begin{equation}
  \label{eq:eRWA2}
e_{\vec{q} \sigma}(t)=\sum_{j=1}^N \sum_{\nu=-1}^1 \beta_j^{\nu} (t) g_q
e^{-i \vec{q} \cdot \vec{r}_j}
\left( \hat{d}_{ g \nu}\cdot \vec{\epsilon}_{\vec{q} \sigma}\right)
\int\limits_0^t d\tau e^{i(\omega_q-\omega_0)\tau}.
\end{equation}
Substituting the last equation into Eq.~(\ref{eq:betaRWA}), we obtain a closed set of equations for $\beta_l^{\eta}$:
\begin{equation}
  \label{eq:betaRWA2}
\dot{\beta}_l^{\eta}=
-\sum_{j=1}^N \sum_{\nu=-1}^1 \sum_{\vec{k},\lambda} \beta_j^{\nu} g_k^2 e^{i \vec{k} \cdot (\vec{r}_l-\vec{r}_j)} \left( \hat{d}_{ \eta g} \cdot \vec{\epsilon}_{\vec{k} \lambda}\right)
\left( \hat{d}_{g \nu} \cdot \vec{\epsilon}_{\vec{k} \lambda}\right) \int_0^t d\tau e^{-i (\omega_0-\omega_k)(\tau-t)}
\end{equation}

Since the polarizations of the photons enter into Eq.~(\ref{eq:betaRWA2}) as separate scalar products, we directly perform the sum over the polarizations using \cite{Smith91}
\begin{equation}
  \label{eq:polarizationRecipe}
\sum_{\lambda}
\left( \hat{d}_{ \eta g} \cdot \vec{\epsilon}_{\vec{k} \lambda}\right)
\left( \hat{d}_{g \nu} \cdot \vec{\epsilon}_{\vec{k} \lambda}\right)=
\hat{d}_{ \eta g} \cdot
\left( \overleftrightarrow{I}-\hat{k} \hat{k}\right)\cdot
\hat{d}_{g \nu}.
\end{equation}
After introducing the short hand notations $\vec{R}_{lj}=\vec{r}_l-\vec{r}_j$, $C_{ \eta \nu}=\hat{d}_{ \eta g} \cdot
\left( \overleftrightarrow{I}-\hat{k} \hat{k}\right)\cdot
\hat{d}_{g \nu}$ with a unit tensor $\overleftrightarrow{I}$,
the resulting set of equations for the atomic excited state amplitudes becomes
\begin{eqnarray}
  \label{eq:betaRWA3}
\eqalign{
\dot{\beta}_l^{\eta}=
-\beta_l^{\eta} \sum_{\vec{k}}  g_k^2   C_{\eta \eta} \int_0^t d\tau e^{-i (\omega_0-\omega_k)(\tau-t)} -\\
-\sum_{j, \nu} \sum_{\vec{k}} \beta_j^{\nu} g_k^2  \left(1-\delta_{l,j}\delta_{\nu,\eta} \right) e^{i \vec{k} \cdot \vec{R}_{l j}} C_{\eta \nu}  \int_0^t d\tau e^{-i (\omega_0-\omega_k)(\tau-t)},
}
\end{eqnarray}
where we have explicitly separated the terms with $j=l$ and $\nu=\eta$ since these terms have different physical interpretations.

The next step in our derivation involves evaluation of the sum over field propagation directions. For this we pass to the free space continuum using the standard form $\sum_{\vec{k}}\rightarrow \frac{V}{(2 \pi c)^3} \int_0^{\infty}d\omega_k \omega_k^2\int d\Omega(\hat{k})$ \cite{Milonni74}, and we arrive at
\begin{equation}
  \label{eq:betaRWA4}
\eqalign{
\dot{\beta}_l^{\eta}=
-\frac{V}{(2 \pi c)^3} \left[
\beta_l^{\eta} \int d\Omega(\hat{k}) C_{\eta \eta} \int_0^{\infty}d\omega_k \omega_k^2 g_k^2
\int_0^t d\tau e^{-i (\omega_0-\omega_k)(\tau-t)} \right. +\\ 
\fl + \left. \sum_{j, \nu}  \beta_j^{\nu} (1-\delta_{l j}\delta_{\nu \eta}) \int d\Omega(\hat{k}) C_{\eta \nu}  \int_0^{\infty}d\omega_k \omega_k^2 g_k^2 e^{i \vec{k}\cdot \vec{R}_{l j}}
 \int_0^t d\tau e^{-i (\omega_0-\omega_k)(\tau-t)}
\right]
}.
\end{equation}

The angular integration can now be performed noting that $\sum_{\vec{k}}e^{-i \vec{k}\cdot \vec{R}}f(k)=\sum_{\vec{k}}e^{i \vec{k}\cdot \vec{R}}f(k)$, $ \int d\Omega(\hat{k}) \left(\overleftrightarrow{I}-\hat{k} \hat{k} \right)=4\pi\frac{2}{3} \overleftrightarrow{I}$
 and $ \int d\Omega(\hat{k}) e^{-i \vec{k}\cdot \vec{R}} \left(\overleftrightarrow{I}-\hat{k} \hat{k} \right)=4\pi \overleftrightarrow{\tau}(k R)$ with $k=|\vec{k}|$, $R_{l j}=|\vec{R}_{l j}|$ and the second rank tensor $\overleftrightarrow{\tau}(k R)$ defined as \cite{Smith91}
\begin{equation}
  \label{eq:tau}
\overleftrightarrow{\tau}(k R)=\left[\overleftrightarrow{I}-\hat{R} \hat{R} \right]\frac{\sin(k R)}{k R}
+\left[\overleftrightarrow{I}-3\hat{R} \hat{R} \right]\left(\frac{\cos(k R)}{k^2 R^2}-\frac{\sin(k R)}{k^3 R^3}\right).
\end{equation}

This results in a set of equations
\begin{equation}
  \label{eq:betaRWA5}
\eqalign{
\dot{\beta}_l^{\eta}=
-\frac{8 \pi^2 d_{e g}^2}{(2 \pi c)^3 \hbar}  \left[
\frac{2}{3}\beta_l^{\eta} \int_0^{\infty}d\omega_k \omega_k^3
\int_0^t d\tau e^{-i (\omega_0-\omega_k)(\tau-t)} +\right.  \\
+ \left .\sum_{j, \nu}  \beta_j^{\nu} (1-\delta_{l j}\delta_{\nu \eta})  
\times \int_0^{\infty}d\omega_k \omega_k^3 D_{\eta \nu}^{l j}
 \int_0^t d\tau e^{-i (\omega_0-\omega_k)(\tau-t)}
\right]
}
\end{equation}
with $D_{\eta \nu}^{l j}=\hat{d}_{\eta g} \cdot \overleftrightarrow{\tau}(k R_{l j}) \cdot \hat{d}_{g \nu}$.

We now turn to the integral over time using the standard expression \cite{Smith91,Guo94}
\begin{equation}
  \label{eq:timeRecipe}
\int_0^t d\tau e^{ i (\omega_k \mp \omega_0)(\tau-t)}=\pi \delta(\omega_k \mp \omega_0)- i P\frac{1}{\omega_k \mp \omega_0},
\end{equation}
where $P$ denotes the principal value.

The set of Eq.~(\ref{eq:betaRWA5}) are now in the form of linear coupled equations for the excitation amplitudes of the different atoms
\begin{eqnarray}
  \label{eq:betaRWA6}
\dot{\beta}_l^{\eta}=-\frac{\Gamma}{2}\beta_l^{\eta}+ i  \left(\frac{\Gamma}{2 \omega_0^3}A \right)\beta_l^{\eta}-\sum_{j=1}^n \sum_{\nu=-1}^1 \frac{3}{2}  \frac{\Gamma}{2 \omega_0^3} \left( \pi B_{l j}^{\eta \nu}-i G_{l j}^{\eta \nu}\right) \beta_j^{\nu}.
\end{eqnarray}
We have explicitly grouped terms where  $\beta_l^{\eta}$ couples to itself and where it couples to different $\beta_j^{\nu}$ as they have different physical interpretations and consequences.

In particular, we recover $\Gamma=\frac{4 d_{eg}^2 k_0^3}{3\hbar}$, the single atom decay rate, in the first term in Eq.~(\ref{eq:betaRWA6}). The second term in Eq.~(\ref{eq:betaRWA6}) with
\begin{equation}
  \label{eq:ARWA}
A=P\int_0^\infty\frac{d\omega_k \omega_k^3}{\omega_k-\omega_0}
\end{equation}
has imaginary value and does not depend on the location or number of the atoms in the ensemble. This terms represents the single atom Lamb shift.
It formally diverges, also when the RWA is not applied, but since it is a constant, we assume that it can be dealt with by the conventional assumption that it is properly included in the measured atomic transition frequency.

Furthermore, in the second line of Eq.~(\ref{eq:betaRWA6}) we have separated the contributions which explicitly depend on the geometry of the atomic sample. The third term is real valued and describes the modification of the decay rate of the atom $l$ much in the analogy to the single atom effects. Unlike the single atom decay, the third term stems from the coupling to all atoms via the field modes and, hence, describes the collective effect:
\begin{eqnarray}
  \label{eq:BRWA}
B_{l j}^{\eta \nu}=(1-\delta_{j l}\delta_{\eta \nu})\int_0^\infty d\omega_k \omega_k^3 \delta(\omega_k-\omega_0) \left(\hat{d}_{\eta g} \cdot \overleftrightarrow{\tau}\left(k R_{l j}\right)\cdot \hat{d}_{g \nu}\right).
\end{eqnarray}
By carrying out the $\omega_k$ integration and using
\begin{eqnarray}
  \label{eq:delta}
(1-\delta_{j l}\delta_{\eta \nu})\left(\hat{d}_{\eta g} \cdot \overleftrightarrow{\tau}\left(k R_{l j}\right)\cdot \hat{d}_{g \nu}\right)=(1-\delta_{j l})\left(\hat{d}_{\eta g} \cdot \overleftrightarrow{\tau}\left(k R_{l j}\right)\cdot \hat{d}_{g \nu}\right),
\end{eqnarray}
due to the orthogonality of $\hat{d}_{\nu g}$ for different Zeeman transitions, Eq.(\ref{eq:BRWA}) can be simplified to the expression
\begin{equation}
  \label{eq:B2}
B_{l j}^{\eta \nu}=(1-\delta_{j l}) \omega_0^3 \left(\hat{d}_{\eta g} \cdot \overleftrightarrow{\tau}\left(k_0 R_{l j}\right)\cdot \hat{d}_{g \nu}\right),
\end{equation}
where $\overleftrightarrow{\tau}$ is given by Eq.~(\ref{eq:tau}) and $k_0=\omega_0/c$.

Finally, the last term of the second line of Eq.~(\ref{eq:betaRWA6}) with imaginary coupling coefficients
\begin{eqnarray}
  \label{eq:G3RWA}
G_{l j}^{\eta \nu}=(1-\delta_{j l})  P \int_{0}^{\infty} \frac{d\omega_k \omega_k^3 }{ \omega_k-\omega_0}
 \left(\hat{d}_{\eta g} \cdot \overleftrightarrow{\tau}\left(k R_{l j}\right)\cdot \hat{d}_{g \nu}\right)
\end{eqnarray}
describe effective level shift of the state $\ket{e_l^{\eta}}$ caused by the dipole coupling to all the other  atoms, mediated by the quantized radiation field.\\

It has been observed in the case of scalar and vector photons interacting with an ensemble of two level atoms, that the RWA Hamiltonian yields an incorrect description of the single atom Lamb shift and the collective coupling of the atoms \cite{Svidzinsky10, Milonni74}. The discrepancy stems from neglecting the contributions from the electric dipole coupling terms which are present in Eq.~(\ref{eq:HV}), but suppressed in the RWA interaction Hamiltonian Eq.~(\ref{eq:HVRWA}) and which drive the off-resonant, virtual transitions depicted schematically with the diagonal dashed arrow in Fig.~\ref{fig:RWA}.

For the case of two-level atoms it was observed that these contributions to the collective shifts can be correctly represented by retaining only the RWA Hamiltonian, if one extends, rather artificially,  the lower integration limit of $\omega_k$ to "$-\infty$" when we go to the free space continuum in Eq.~(\ref{eq:betaRWA4}) \cite{Svidzinsky10, Milonni74}. Using this rule, the coefficients in our equations of motion for the case of multilevel atoms change. The single atom Lamb shift now becomes
\begin{equation}
  \label{eq:ARWAAnsatz}
A^{(A)}=P\int_{-\infty}^\infty\frac{d\omega_k \omega_k^3}{\omega_k-\omega_0}.
\end{equation}
The decay rate contributions $B_j^{\nu}$ due to the collective effects described by Eq.~(\ref{eq:BRWA}) are unchanged, since $\omega_0$ is positive according to our definition. In contrast, the collective level shifts of state $\ket{e_l^{\eta}}$ due to the presence of all other $N-1$ atoms change to
\begin{eqnarray}
  \label{eq:G3RWAAnsatz}
G_{l j}^{\eta \nu(A)}=(1-\delta_{j l})  P \int_{-\infty}^{\infty} \frac{d\omega_k \omega_k^3 }{ \omega_k-\omega_0}
 \left(\hat{d}_{\eta g}^l \cdot \overleftrightarrow{\tau}\left(k R_{l j}\right)\cdot \hat{d}_{g \nu}^j\right).
\end{eqnarray}

The equations of motion Eq.~(\ref{eq:betaRWA6}) with the collective interaction coefficients in the form Eq.~(\ref{eq:BRWA}) and Eq.~(\ref{eq:G3RWAAnsatz}) are consistent with the result by Friedberg \textit{et al.} \cite{Friedberg08b} based on the direct implementation of the pairwise dipole-dipole interaction between atoms. Since the mere observation of the equivalence between the two results does not offer a solid physical argument that they should also be equivalent with the result of a full treatment of the interaction Hamiltonian, we will in the following section derive the equations of motion of the atomic excited state amplitudes taking into account the off-resonant transitions caused by the non-RWA terms in the interaction Hamiltonian Eq.~(\ref{eq:HV_general}).

\section{Beyond the Rotating Wave Approximation}
\label{sec:full}
The derivation in the present section to a large extent follows the derivation in Sec.~\ref{sec:RWA}, and we will explicitly highlight the additional terms that appear due to the virtual transitions to states with more than a single excitation. Due to the parity conservation dictated by $H_V$, the only state coupled directly to $\ket{e_j^{\nu}}\ket{0}$ is $ \ket{e_n^{\nu} e_m^{\mu}}\ket{1_{\vec{k}\lambda}}$ with two atoms excited and one photon present in the electromagnetic field. Due to its large violation of energy conservation this transition is significantly suppressed relative to the energy conserving transition to $\ket{g}\ket{1_{\vec{k} \lambda}}$, and the coupling to states with even higher excitations via the state $ \ket{e_n^{\nu} e_m^{\mu}}\ket{1_{\vec{k}\lambda}}$ can be safely neglected. The state vector of the atoms and the quantized field can hence be expanded on the form \cite{Stephen64,Svidzinsky10}
\begin{equation}
\label{eq:psi}
\eqalign{
 \ket{\psi(t)}=\sum_{j=1}^N \sum_{\nu=-1}^1 \beta_j^{\nu}(t)e^{-i\omega_0 t}\ket{e_j^{\nu}}\ket{0}
+\sum_{\vec{k},\lambda} e_{\vec{k} \lambda}(t)e^{-i \omega_k t} \ket{g}\ket{1_{\vec{k} \lambda}}+\\
+\sum_{n=1  }^N\sum_{m=n+1}^N \sum_{\nu, \mu=-1}^1 \sum_{\vec{k},\lambda}\alpha_{m n,\vec{k}\lambda}^{\nu \mu} (t) e^{-i (\omega_k+2 \omega_0)t} \ket{e_n^{\nu} e_m^{\mu}}\ket{1_{\vec{k}\lambda}},
}
\end{equation}
where in addition to Eq.~(\ref{eq:psiRWA}) we have introduced a state $\ket{e_n^{\nu} e_m^{\mu}}\ket{1_{\vec{k}\lambda}}=\ket{g_1,...,e_n^{\nu},...,e_m^{\mu},...,g_N}\ket{1_{\vec{k}\lambda}}$ with  two atoms excited and one photon in the field mode.

The substitution of Eq.~(\ref{eq:psi}) into the time dependent Schr\"odinger equation with the Hamiltonian from Eq.~(\ref{eq:H}) and Eq.~(\ref{eq:HV}) yields:
\begin{equation}
  \label{eq:beta}
\eqalign{
 \dot{\beta}_{l}^{\eta}=-\sum_{\vec{k}, \lambda} e_{\vec{k} \lambda}g_k e^{-i((\omega_k-\omega_0)t- \vec{k} \cdot \vec{r}_l)}\left( \hat{d}_{\eta g}\cdot \vec{\epsilon}_{\vec{k} \lambda}\right)-\\
- \sum_{n=1}^N \sum_{m=n+1  }^N \sum_{\nu, \mu=-1}^1 \sum_{\vec{k},\lambda} \alpha_{m n,\vec{k} \lambda}^{\nu \mu} g_k e^{-i(\omega_k+\omega_0)t} \times \\
\times \left[e^{i \vec{k} \cdot \vec{r}_m} \left( \hat{d}_{g \mu}\cdot \vec{\epsilon}_{\vec{k} \lambda}\right) \delta_{l n} \delta_{\eta \nu} +
e^{i \vec{k} \cdot \vec{r}_n} \left( \hat{d}_{g \nu}\cdot \vec{\epsilon}_{\vec{k} \lambda}\right) \delta_{l m} \delta_{\eta \mu}
\right],
}
\end{equation}
\begin{equation}
  \label{eq:e}
 \dot{e}_{\vec{q} \sigma}=\sum_{j=1}^N \sum_{\nu=-1}^1 \beta_j^{\nu}g_q e^{i((\omega_q-\omega_0)t- \vec{q} \cdot \vec{r}_j)}\left( \hat{d}_{g \nu}\cdot \vec{\epsilon}_{\vec{q} \sigma}\right)
\end{equation}
and
\begin{equation}
  \label{eq:alpha}
\eqalign{
 \dot{\alpha}_{p r,\vec{q} \sigma}^{\rho \pi}=\beta_r^{\rho}g_q e^{i((\omega_q+\omega_0)t- \vec{q} \cdot \vec{r}_p)}\left( \hat{d}_{\pi g}\cdot \vec{\epsilon}_{\vec{q} \sigma}\right)+\beta_p^{\pi}g_q e^{i((\omega_q+\omega_0)t- \vec{q} \cdot \vec{r}_r)}
\left(\hat{d}_{\rho g}\cdot \vec{\epsilon}_{\vec{q} \sigma}\right)
}
\end{equation}

The coupling to the atomic ground state $\ket{g}$, which has exactly the same form as in the RWA case, and the coupling to states with two excited atoms are obtained using the fact that there are only four nonzero matrix elements with the Hamiltonian Eq.~(\ref{eq:HV}):
$\bra{0}\bra{e_l^{\eta}} H_{V}\ket{g}\ket{1_{\vec{k} \lambda}}$, $\bra{1_{\vec{k} \lambda}}\bra{g} H_{V}\ket {e_l^{\eta}}\ket{0}$, $\bra{0}\bra{e_l^{\eta}} H_{V}\ket{e_n^{\nu} e_m^{\mu}}\ket{1_{\vec{k} \lambda}}$ and $\bra{1_{\vec{k} \lambda}}\bra{e_m^{\mu}e_n^{\nu}} H_{V}\ket{e_l^{\eta}}\ket{0}$. We present the explicit form of these matrix elements in the \ref{sec:AppendixA}.

Since the mathematical structure of the new terms in Eq.~(\ref{eq:beta})-(\ref{eq:alpha}) is formally similar to the corresponding terms for excitation number conserving transitions, we  integrate Eq.~(\ref{eq:e})-(\ref{eq:alpha}) using the initial conditions $e_{\vec{q} \sigma}(0)=0$ and $\alpha_{p r,\vec{q} \sigma}^{\rho \pi}(0)=0$, as in Sec.~\ref{sec:RWA}. Here again we additionally assume the Markovian approximation, and obtain
\begin{equation}
  \label{eq:6e}
\eqalign{
e_{\vec{q} \sigma}(t)=\sum_{j=1}^N \sum_{\nu=-1}^1 \beta_j^{\nu} (t) g_q
e^{-i \vec{q} \cdot \vec{r}_j}
\left( \hat{d}_{ g \nu}\cdot \vec{\epsilon}_{\vec{q} \sigma}\right)
\int\limits_0^t d\tau e^{i(\omega_q-\omega_0)\tau}
}
\end{equation}
and
\begin{equation}
  \label{eq:7alpha}
\fl
\eqalign{
\alpha_{p r,\vec{q} \sigma}^{\rho \pi}(t)
=\left[\beta_r^{\rho}(t)
e^{-i \vec{q} \cdot \vec{r}_p}
\left( \hat{d}_{ \pi g}\cdot \vec{\epsilon}_{\vec{q} \sigma}\right)  +\beta_p^{\pi} (t)
e^{-i \vec{q} \cdot \vec{r}_r}
\left( \hat{d}_{ \rho g}\cdot \vec{\epsilon}_{\vec{q} \sigma} \right)
\right]  g_q  \int\limits_0^t d\tau
 e^{i(\omega_q+\omega_0)\tau}.
}
\end{equation}
These expressions are inserted into Eq.~(\ref{eq:beta}) to yield a closed set of equations for the single-atom-excitation amplitudes $\beta_l^{\eta}$:
\begin{equation}
  \label{eq:8beta}
\eqalign{
\fl
\dot{\beta}_l^{\eta}=
-\sum_{j=1}^N \sum_{\nu=-1}^1 \sum_{\vec{k},\lambda} \beta_j^{\nu} g_k^2 e^{i \vec{k} \cdot (\vec{r}_l-\vec{r}_j)} \left( \hat{d}_{ \eta g} \cdot \vec{\epsilon}_{\vec{k} \lambda}\right)
\left( \hat{d}_{g \nu} \cdot \vec{\epsilon}_{\vec{k} \lambda}\right) \int_0^t d\tau e^{-i (\omega_0-\omega_k)(\tau-t)}-\\
-\sum_{n=1}^N \sum_{m=n+1}^N\sum_{\nu,\mu=-1}^1 \sum_{\vec{k},\lambda}  g_k^2 \int_0^t d\tau e^{-i (\omega_0+\omega_k)(\tau-t)}\times \\
\fl
\times \left[ \right.
\beta_n^{\nu}
\left( \hat{d}_{g \mu} \cdot \vec{\epsilon}_{\vec{k} \lambda}\right)
\left( \hat{d}_{ \mu g} \cdot \vec{\epsilon}_{\vec{k} \lambda}\right)
\delta_{l n} \delta_{\eta \nu}+
\beta_m^{\mu}
\left( \hat{d}_{g \mu} \cdot \vec{\epsilon}_{\vec{k} \lambda}\right)
\left( \hat{d}_{ \nu g} \cdot \vec{\epsilon}_{\vec{k} \lambda}\right)
e^{i\vec{k}\cdot (\vec{r}_m-\vec{r}_n)}
\delta_{l n} \delta_{\eta \nu}+\\
\fl
+\beta_n^{\nu}
\left( \hat{d}_{g \nu} \cdot \vec{\epsilon}_{\vec{k} \lambda}\right)
\left( \hat{d}_{ \mu g} \cdot \vec{\epsilon}_{\vec{k} \lambda}\right)
e^{i\vec{k}\cdot (\vec{r}_n-\vec{r}_m)}
\delta_{l m} \delta_{\eta \mu}+\beta_m^{\mu}
\left( \hat{d}_{g \nu} \cdot \vec{\epsilon}_{\vec{k} \lambda}\right)
\left( \hat{d}_{ \nu g} \cdot \vec{\epsilon}_{\vec{k} \lambda}\right)
\delta_{l m} \delta_{\eta \mu}
\left. \right].
}
\end{equation}
The first coupling term is equivalent to the energy conserving contributions  in Eq.~(\ref{eq:betaRWA}) in the RWA, while the subsequent terms come from the coupling to states with different excitation numbers. Even though the population of these states is negligible due to their large violation of energy conservation, they perturb the coupling between the states with a single atomic excitation. In order to interpret their physical meaning we perform a few mathematical manipulations of the non-RWA coupling terms. Since the polarizations of photons enter into Eq.~(\ref{eq:8beta}) as separate scalar products, we directly perform polarization summations according to Eq.~(\ref{eq:polarizationRecipe}). After introducing short hand notation $\theta_{n m}=0$ if $m\le n$ and $\theta_{n m}=1$ if $m > n$, the  set of equations simplifies
\begin{equation}
  \label{eq:beta2}
\eqalign{
\dot{\beta}_l^{\eta}=
-\sum_{j=1}^N \sum_{\nu=-1}^1 \sum_{\vec{k}} \beta_j^{\nu} g_k^2  \int_0^t d\tau e^{-i (\omega_0-\omega_k)(\tau-t)} e^{i \vec{k} \cdot \vec{R}_{l j}} C_{\eta \nu}-\\
\fl
-\sum_{m=1}^N \sum_{\mu=-1}^1 \sum_{\vec{k}}  g_k^2   [
\beta_l^{\eta} C_{\mu \mu}( \theta_{l m}+\theta_{m l})
+\beta_m^{\mu} e^{i \vec{k} \cdot \vec{R}_{m l}} C_{\eta \mu} (\theta_{l m}+ \theta_{m l})
]\int_0^t d\tau e^{-i (\omega_0+\omega_k)(\tau-t)}.
}
\end{equation}
Regrouping the terms and noting that $\theta_{l j}+\theta_{j l}=1-\delta_{l j}$ we expand Eq.~(\ref{eq:beta2}):
\begin{equation}
  \label{eq:10beta}
\eqalign{
\dot{\beta}_l^{\eta}=
- \beta_l^{\eta}  \sum_{\vec{k}}  g_k^2  \left [C_{\eta \eta} \int_0^t d\tau e^{-i (\omega_0-\omega_k)(\tau-t)}+\right.\\
+ \sum_{j, \nu}   C_{\nu \nu} (1-\delta_{l j}) \left(1-\delta_{l,j}\delta_{\nu,\eta} \right)
\left.  \int_0^t d\tau e^{-i (\omega_0+\omega_k)(\tau-t)}\right]-\\
-\sum_{j, \nu} \sum_{\vec{k}} \beta_j^{\nu} g_k^2  \left(1-\delta_{l,j}\delta_{\nu,\eta} \right) C_{\eta \nu}  \times  \\
\times  \left[ e^{i \vec{k} \cdot \vec{R}_{l j}} \int_0^t d\tau e^{-i (\omega_0-\omega_k)(\tau-t)}+(1-\delta_{l j})  e^{i \vec{k} \cdot \vec{R}_{j l}}  \int_0^t d\tau e^{-i (\omega_0+\omega_k)(\tau-t)}\right],
}
\end{equation}
where we have explicitly separated the terms with $j=l$ and $\nu=\eta$ to write the equations in a form similar to Eq.~(\ref{eq:betaRWA3}).

The first term in the equation refers to single atom effects, i.e., the real part yields the decay and the imaginary part yields the Lamb shift of the atomic excited state. Compared with the RWA analysis, which yields the first coefficient in the square bracket, the coupling to the two-excitation sector provides a correction to the single atom coupling term. The second term couples the excited state amplitudes of the different atoms to each other and is responsible for the collective effects. Also here, the first term in the square bracket is obtained in the RWA, while the second term is the non-RWA correction to the collective coupling terms. As in the previous section, we can introduce the free space continuum of field modes as in Eq.~(\ref{eq:betaRWA4}) and perform the angular integration as in Eq.~(\ref{eq:betaRWA5}), which yields

\begin{equation}
  \label{eq:beta3}
\fl
\eqalign{
\dot{\beta}_l^{\eta}=
-\frac{8 \pi^2 d_{e g}^2}{(2 \pi c)^3 \hbar} \frac{2}{3}\beta_l^{\eta} \int_0^{\infty}d\omega_k \omega_k^3 \times \\
\times \left[ \int_0^t d\tau e^{-i (\omega_0-\omega_k)(\tau-t)} +  \sum_{j, \nu} (1-\delta_{l j})(1-\delta_{l j}\delta_{\nu \eta})
\int_0^t d\tau e^{-i (\omega_0+\omega_k)(\tau-t)} \right] +\\ \\
+ \sum_{j, \nu}  \beta_j^{\nu} (1-\delta_{l j}\delta_{\nu \eta})
 \int_0^{\infty}d\omega_k \omega_k^3 D_{\eta \nu}^{l j}\times \\
\times \left[ \int_0^t d\tau e^{-i (\omega_0-\omega_k)(\tau-t)} +  (1-\delta_{l j}) \int_0^t d\tau e^{-i (\omega_0+\omega_k)(\tau-t)}
\right].
}
\end{equation}
Finally, carrying out the time integrals and using $\sum_{j=1}^N \sum_{\nu=-1}^1 (1-\delta_{j l})(1-\delta_{j l}\delta_{\eta \nu})=3(N-1)$, we arrive at the equation Eq.~(\ref{eq:betaRWA6}), but with the modified coefficients,
\begin{equation}
  \label{eq:A}
A^{(F)}=P\int_0^\infty\frac{d\omega_k \omega_k^3}{\omega_k-\omega_0}
+3(N-1)P\int_0^\infty\frac{d\omega_k \omega_k^3}{\omega_k+\omega_0}.
\end{equation}
The first term yields the RWA contribution to the Lamb shift of the excited atomic state found also in Eq.~(\ref{eq:ARWA}). The second term is
due to the virtual transition with any one of the other $(N-1)$ ground state atoms being excited, i.e., it represents the single atom ground state Lamb shift of those atoms \cite{Svidzinsky10,Scully09b,Milonni74}.

The coupling of different atomic excited state amplitudes in  Eq.~(\ref{eq:betaRWA6}) explicitly depends on the geometry of the sample. The real part describes the modification of the decay rate of the atom $l$ due to all the other atoms around it
\begin{equation}
  \label{eq:B2}
B_{l j}^{\eta \nu (F)}=(1-\delta_{j l}) \omega_0^3 \left(\hat{d}_{\eta g} \cdot \overleftrightarrow{\tau}\left(k_0 R_{l j}\right)\cdot \hat{d}_{g \nu}\right)
\end{equation}
and is unchanged, relative to the RWA case. The virtual transitions to the higher excited states thus turn out to have no direct influence on the decay terms responsible for the emission of real photons.

On the contrary, the imaginary coupling terms in Eq.~(\ref{eq:betaRWA6}), which yield dispersive coupling among the excited state atomic amplitudes and collective level shift effects, acquires an extra contribution, relative to the RWA case in Eq.~(\ref{eq:G3RWA}),
\begin{equation}
  \label{eq:G}
\eqalign{
G_{l j}^{\eta \nu (F)}=(1-\delta_{j l}\delta_{\eta \nu})  P \int_0^{\infty} \frac{d\omega_k \omega_k^3 }{ \omega_k-\omega_0}
 \left(\hat{d}_{\eta g} \cdot \overleftrightarrow{\tau}\left(k R_{l j}\right)\cdot \hat{d}_{g \nu}\right)+\\ \\
+(1-\delta_{j l}\delta_{\eta \nu})(1-\delta_{j l})  P \int_0^{\infty} \frac{d\omega_k \omega_k^3 }{ \omega_k+\omega_0} \left(\hat{d}_{g \nu} \cdot \overleftrightarrow{\tau}\left(k R_{l j}\right)\cdot \hat{d}_{\eta g}\right).
}
\end{equation}
Noting that from the definition of $\overleftrightarrow{\tau}$ we have $$\hat{d}_{\eta g} \cdot \overleftrightarrow{\tau}\left(k R_{l j}\right)\cdot \hat{d}_{g \nu}=\hat{d}_{g \nu} \cdot \overleftrightarrow{\tau}\left(k R_{l j}\right)\cdot \hat{d}_{\eta g}$$ and using Eq.~(\ref{eq:delta}), we can simplify Eq.~(\ref{eq:G}) and combine the terms with the two frequency denominators
\begin{equation}
  \label{eq:G3}
\eqalign{
G_{l j}^{\eta \nu (F)}=(1-\delta_{j l})  P \int_{-\infty}^{\infty} \frac{d\omega_k \omega_k^3 }{ \omega_k-\omega_0}
 \left(\hat{d}_{\eta g} \cdot \overleftrightarrow{\tau}\left(k R_{l j}\right)\cdot \hat{d}_{g \nu}\right).
}
\end{equation}
 We have thus recovered exactly the same expression as one obtains from the RWA ansatz and the purely formal extension of the frequency integral to negative values in  Sec.~\ref{sec:RWA}. In the \ref{sec:AppendixB} we evaluate the last integral explicitly.

\begin{figure*}[tbp]
  \centering
{\includegraphics{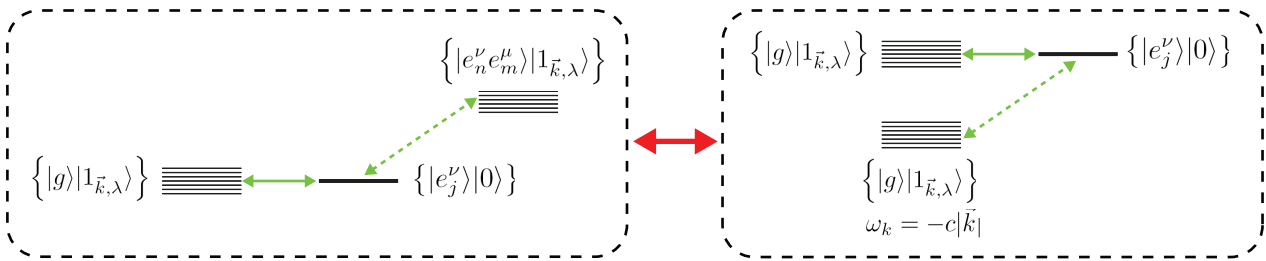}}
  \caption{(Color online) Quantized matter light interaction treated by two different methods. The left panel shows how the full Hamiltonian Eq.~(\ref{eq:HV}), resonantly couples the states $\{ \ket{e_j^{\nu}}\ket{0} \}$ to the states $\{ \ket{g}\ket{\vec{k},\lambda} \}$ and off-resonantly (dashed line) to the higher excited states $\{\ket{e_n^{\nu}e_m^{\mu}}\ket{\vec{k},\lambda} \}$. The right panel shows the same resonant coupling and artificial off-resonant coupling to states with negative photon energy. In this letter we show that the two methods yield identical corrections to the interactions in the resonant manifold of states.
 \label{fig:equivalence}}
\end{figure*}

\section{Conclusion}
\label{sec:Conclusion}
Using the Rotating Wave Approximation and the full Hamiltonian, we have presented parallel derivations of the equations of motion describing the atomic population dynamics in a system of $N$ multilevel atoms interacting with the quantized vector electromagnetic field. For the collective interaction part we find agreement of the full method with the results of the RWA, when the latter is extended by contributions from negative frequency photon states. Note that in contrast to the collective effects, the Lamb shift is not recovered in this method. A similar partial agreement was found for simpler atomic transitions and for scalar electromagnetic fields. From the physical point of view the single atom Lamb shift can be included into the definitions of the atomic levels, and so the collective atomic interaction part becomes the leading contribution to the system dynamics. From the mathematical point of view, the RWA with the simple extension to negative frequencies is easier to handle in practical calculations, and also easier to generalize to more complicated problems. Nevertheless, neither the simple scalar case nor our more general analysis, offer a clear explanation of its validity.

With reference to Fig.~\ref{fig:equivalence}, one may advocate that the off-resonant transitions governed by the non-RWA terms in the full Hamiltonian explore the same frequency range and violation of energy conservation. This is, indeed, what we see explicitly in passing from Eq.~(\ref{eq:G}) to  Eq.~(\ref{eq:G3}). The same figure, however, shows that the two  state spaces coupled by the off-resonant transitions differ substantially by the kind of states involved. In particular these states have very different dimensionality accompanying the combinatorial aspects in selecting two rather than zero  excited atoms. The fact that the off resonant coupling to the excited states in the left part of Fig.~\ref{fig:equivalence} leads to the same reduced dynamics in the single-excitation subspace, as the artificial coupling to the negative energy continuum, must be due to the detailed counting of coupling terms, responsible for the similar equivalence observed by Milonni \textit{et al.} \cite{Milonni74}. The agreement may still be surprising, given the fact that the states coupled experience the excited state Zeeman degeneracy, while the negative frequency continuum involves only the ground state atomic manifold. The rotational invariance of the coupling to all field propagation directions and polarizations, see as well \cite{Kiffner07}, must be at the root of the apparent suppression of any azimuthal quantum number dependence of the result. In an accompanying paper \cite{Miroshnychenko13} we study the case of atoms between two plane mirrors, which impose boundary conditions on the quantized electro-magnetic field and breaks the rotational invariance of the atom-field coupling. In this configuration the same rule with the extension of the photon frequency integrals allows to recover the collective energy shift terms, although the density of photon states is fundamentally different from the free atoms case. While still not providing a simple physical argument, our studies support the possibility, that the RWA Hamiltonian extended to negative frequencies may universally capture the dispersive effects of the coupling to the off-resonant, triply excited states.

\section*{Acknowledgments}
 We acknowledge financial support from the project MALICIA under FET-Open grant number 265522. Y.~M. acknowledges fruitful discussions with Sevilay~Sevincli and Uffe~V.~Poulsen at the early stage of the project.

\appendix
\section{}
\label{sec:AppendixA}
The matrix elements with single atomic excitations are evaluated straightforwardly
\begin{equation}
  \label{eq:matr0eHg1}
\bra{0}\bra{e_l^{\eta}} H_{V}\ket{g}\ket{1_{\vec{k} \lambda}}=-i\hbar g_k
e^{i \vec{k} \cdot \vec{r}_l} \left( \hat{d}_{\eta g} \cdot \vec{\epsilon}_{\vec{k} \lambda} \right)
\end{equation}
and
\begin{equation}
  \label{eq:matr1gHe0}
\bra{1_{\vec{k} \lambda}}\bra{g} H_{V}\ket {e_l^{\eta}}\ket{0}=\left( \bra{0}\bra{e_l^{\eta}} H_{int}\ket{g}\ket{1_{\vec{k} \lambda}}\right)^+.
\end{equation}
The matrix elements with two atoms excited are evaluated after noting that $\vec{\sigma}_{ge}^j\ket{e_n^{\nu} e_m^{\mu}}=0$ if $j\neq n,m$ and otherwise we can formally write $\vec{\sigma}_{ge}^m\ket{e_n^{\nu} e_m^{\mu}}=\ket{e_n^{\nu}} \vec{\sigma}_{ge}^m \ket{e_m^{\mu}}$:
\begin{equation}
  \label{eq:matr0eHee1}
\fl
\eqalign{
\bra{0}\bra{e_l^{\eta}} H_{V}\ket{e_n^{\nu} e_m^{\mu}}\ket{1_{\vec{k} \lambda}}=-i\hbar g_k  \left[ e^{i \vec{k} \cdot \vec{r}_m} \left( \hat{d}_{g \mu}\cdot \vec{\epsilon}_{\vec{k} \lambda} \right)\delta_{l n} \delta_{\eta \nu}
+ e^{i \vec{k} \cdot \vec{r}_n} \left( \hat{d}_{g \nu}\cdot \vec{\epsilon}_{\vec{k} \lambda} \right)\delta_{l m} \delta_{\eta \mu} \right]
}
\end{equation}
and
\begin{equation}
  \label{eq:matr1eeHe0}
\eqalign{
\bra{1_{\vec{k} \lambda}}\bra{e_m^{\mu}e_n^{\nu}} H_{V}\ket{e_l^{\eta}}\ket{0}=\left( \bra{0}\bra{e_l^{\eta}} H_{V}\ket{e_n^{\nu} e_m^{\mu}}\ket{1_{\vec{k} \lambda}}\right)^+.
}
\end{equation}

\section{}
\label{sec:AppendixB}
Since $\overleftrightarrow{\tau}$ given by Eq.~(\ref{eq:tau}) consists  of three parts, we integrate each part separately using \cite{McLone65}
\begin{equation}
  \label{eq:kksin}
P\int_{-\infty}^{\infty}d k \frac{k^2 \sin(k R)}{k-k_0}=\pi k_0^2 \cos(k R),
\end{equation}
\begin{equation}
  \label{eq:kcos}
P\int_{-\infty}^{\infty}d k \frac{k \cos(k R)}{k-k_0}=-\pi k_0 \sin(k R)
\end{equation}
and
\begin{equation}
  \label{eq:sin}
P\int_{-\infty}^{\infty}d k \frac{ \sin(k R)}{k-k_0}=\pi  \cos(k R)
\end{equation}
with $k=\omega_k/c$ and $k_0=\omega_0/c$.
Consequently, defining
\begin{equation}
  \label{eq:gamma}
\overleftrightarrow{\gamma}(k R)=\left[\overleftrightarrow{I}-\hat{R} \hat{R} \right]\frac{\cos(k R)}{k R}+\left[\overleftrightarrow{I}-3\hat{R} \hat{R} \right]\left(\frac{\sin(k R)}{k^2 R^2}+\frac{\cos(k R)}{k^3 R^3}\right),
\end{equation}
Eq.~(\ref{eq:G3}) can be rewritten as
\begin{equation}
  \label{eq:G4}
G_{l j}^{\nu \eta (F)}=(1-\delta_{j l})\left(\hat{d}_{\eta g} \cdot \overleftrightarrow{\gamma}\left(k R_{l j}\right)\cdot \hat{d}_{g \nu}\right).
\end{equation}


\end{document}